\documentclass[preprint]{aastex}
\usepackage{emulateapj5}
\usepackage{epsfig}
 
\begin{document}

\title{
Growth of a peanut shaped bulge via resonant trapping of stellar
orbits in the vertical Inner Lindblad Resonances
}

\author{ A.~C. Quillen\altaffilmark{1} 
}
\affil{Department of Physics and Astronomy, University of Rochester, Rochester, NY 14627; aquillen@pas.rochester.edu}
\altaffiltext{1}{
Visitor, Physics Department, Technion, Israel Institute of Technology}

\begin{abstract}             
We present a simple resonant Hamiltonian model for the vertical
response of a stellar disk to the growth of a bar perturbation.
As a bar perturbation grows stars
become trapped in vertical Inner  Lindblad resonances and are
lifted into higher amplitude orbits.
The vertical structure of a
boxy and peanut shaped bulge as a function of radius 
and azimuthal angle in the galaxy plane can be predicted
from the strength and speed of the bar perturbation and
the derivatives of the gravitational potential.
This model predicts that stars on the outer 
side of the resonance are lifted higher than stars
on the inner side, offering an explanation for the sharp
outer edge of the boxy/peanut.
\end{abstract}             


\section{Introduction}   

The existence of spiral galaxy bulges with boxy or peanut
shapes has been recognized for many years
through the study of edge-on galaxies
(e.g., \citealt{burbridge,devauc}).
Recent work has found that edge-on galaxies with
prominent boxy or peanuts shaped bulges in most cases show kinematic evidence
for non-circular motions  associated with a bar,
whereas galaxies lacking boxy or peanut shaped bulges
do not show kinematic evidence for a bar
\citep{bureau,merrifield}. 
A few cases exist where highly inclined
systems exhibit both bars and peanuts (e.g., NGC 7582, \citealt{quillen}
and NGC~4442, \citealt{bettoni}).
The link between a galactic bar and a boxy/peanut shaped bulge has
been firmly established.

Mechanisms proposed to explain the formation of boxy and peanut shaped
structures include accretion of small
satellite galaxies \citep{petrou}, bar-buckling
via the fire-hose type instability
\citep{toomre,raha,merritt,fridman} 
and resonant heating \citep{pfenniger84,pfenniger85,combes}.
The fire-hose (bar-buckling) instability is a global time dependent
instability which relies on the centrifugal
force caused by highly eccentric orbits in the plane
caused by the bar and is damped via a dissipative process similar
to Landau damping \citep{fridman}.    The resonant heating model
has been inspired by the study and classification of orbit families 
in 3-dimensional systems \citep{pfenniger91} and through
the phenomena displayed in N-body simulations \citep{combes}.

No study has yet considered 
the evolution of 3 dimensional stellar orbits in a barred system that is
varying with time.
In this paper we investigate the possibility that the growth
of the bar itself traps particles in the vertical 
Inner Lindblad Resonances (ILRs).  
We develop a simple Hamiltonian analytical model for these resonances
and use it to predict the vertical distribution
of a bar following the growth of the bar.
Our approach of considering resonance trapping
is similar to that explored by \cite{sridhar}
for disk heating.  However their model is for the general
case of a growing axisymmetric disk, whereas our
model applies specifically to a growing bar perturbation.

\section{Dynamical Model}

To exihibit resonant phenomena 
\citet{cont75} showed that the Hamiltonian must be expanded
to at least third order in the epicyclic approximation.
When this is done there are higher order cross terms in the action variables
and the model will exhibit phenomenology
such as the bifurcation of orbit families seen in  characteristic
diagrams which are created through numerical integration of orbits.

We first consider the Hamiltonian lacking non-axisymmetric or time dependent perturbations
\begin{equation}
H_0(L,p_r,p_z;\theta,r,z) 
= {L^2\over  2 r^2} + {p_r^2\over 2} + {p_z^2 \over 2} + V_0(r,z)
\end{equation}
where $V_0$ is the gravitational potential which is assumed
to be axisymmetric and $L,p_r,p_z$ are the momenta
conjugate to the cylindrical coordinates $\theta,r,z$.
For the 2 dimensional problem restricted to the Galactic plane
($z=0,p_z=0$),
\citet{cont75} showed how to put the Hamiltonian in the form 
\begin{equation}
K_0(I_1,I_2;\theta_1,\theta_2)  
= h + \kappa I_1 + \Omega I_2 + a I_1^2 + 2 b I_1 I_2 + c I_2^2 ... 
\end{equation}
in a third order epicyclic approximation.
The action variables $I_1 = {1\over 2 \pi} \int \dot r dr$ and $I_2 = J_0 - J_c$
are integrals of motion. 
$J_0$ is the particle's angular momentum, and $J_c$ is the angular
momentum of a particle in a circular orbit at a radius $r_c$
which is the radius of a circular orbit with energy $h$.
$\theta_1$ is the epicyclic angle and $\theta_2$ is
the azimuth of the epicyclic center.  $\theta_1$ and $\theta_2$
are the angle variables conjugate to $I_1$ and $I_2$.
$\kappa$ is the epicyclic frequency and $\Omega$ the angular
rotation rate; both are evaluated at $r=r_c,z=0$.

To extend the theory into the third dimension we must
add additional terms to the restricted Hamiltonian which depend upon $z$;
\begin{equation}
H_0 = K_0 + H_{0z}.
\end{equation}
We expand the potential $V_0$  about $z$ and $r_c$
defining the variables 
\begin{eqnarray}
\nu^2   &=& {\partial^2 V_0  \over \partial z^2} \\
\lambda &=& 
  {\partial^4 V_0    \over \partial z^4} \nonumber 
\end{eqnarray}
which are evaluated at $z=0,r=r_c$.
Inserting this into the full Hamiltonian we find
\begin{eqnarray}
\label{Hz}
H_{0z}(&L,p_r,p_z;\theta,r,z) = 
{1\over 2} (p_z^2 + \nu^2 z^2) + {\lambda  z^4 \over 4!} 
~~~~~~~~~~~~~~~~~~~~ \\
&+ {\nu^2 z^2 \over 2} \left[{ 
                       r_c \Omega^2 (r- r_c) 
      + {1\over 2}(\kappa^2 - 3 \Omega^2) (r- r_c)^2 }\right] \nonumber
\end{eqnarray}
where we have made use of the relations
$\Omega^2 = {1\over r} {\partial V_0 \over  \partial r}$,
$\kappa^2 = {\partial^2 V_0 \over  \partial r^2} + 3 \Omega^2$
and both $\Omega$ and $\kappa$ are evaluated at $r=r_c,z=0$.

For this Hamiltonian following \citet{cont75} we can choose a third action
variable $I_3$ to represent the amplitude of vertical  oscillations.
\begin{eqnarray}
\label{zdef}
z &\approx& \left({2 I_3  \over \nu}\right)^{1/2} \cos(\theta_3) \\
{d z \over d t} &\approx& 
  -\left({2 I_3  \nu}\right)^{1/2} \sin(\theta_3) \nonumber
\end{eqnarray}
where $\nu$ is the vertical oscillation frequency and $\theta_3$
is the angle associated with the vertical oscillations.
The above expressions are given to first order in $I_3^{1/2}$ and 
require higher order terms to be exact.

After a couple of canonical transformations (see appendix)
the Hamiltonian to second order in $I_3^{1/2}$ takes the form 
\begin{equation}
\label{Hz_can}
H_{0z}(I_1,I_2,I_3;\theta_1,\theta_2,\theta_3)  = 
\nu I_3 + {\lambda I_3^2  \over 32  \nu^2}
+ (\kappa^2 - 3 \Omega^2) {\nu I_1 I_3 \over 4\kappa }.
\end{equation}
Since this is independent of the angles $\theta_1,\theta_2,\theta_3$,
the momenta
$I_1,I_2,I_3$ are conserved quantities and
we say we have put the non-perturbed Hamiltonian, (Eqn.~\ref{Hz}),
in action angle form.

\subsection{Perturbation}

We now consider the form of the perturbation of the gravitational
potential caused by a bar.
In the plane of the galaxy we can expand the gravitational
potential in terms of Fourier components 
\begin{equation}
V_1(r,\theta,z=0) = \sum_m A_m \cos(m (\theta - \Omega_b t))
\end{equation}
(for example see \citealt{n4314}).
The strongest term due to the bar should be the $m=2$
term with $A_2 < 0$ so that the bar major axis
lies along $\theta - \Omega_b t = 0, \pi.$
For small $z$ we assume that the potential can be expanded 
so that we can write for each $m$ Fourier component
\begin{equation}
V_{1m}(r,\theta,z) =  A_m \cos(m (\theta - \Omega_b t)) 
\left(1 - {z^2 \over 2h_{1}^2}\right)
\end{equation}
where $h_{1}$ is the scale height of the perturbation.
We also assume that $A_m$ and $h_{1}$ 
vary only weakly with radius.
Since $z^2 \approx {2 I_3\over \nu} \cos^2(\theta_3)$
we can rewrite the perturbation components to second order in $I_3^{1/2}$
\begin{equation}
V_{1m}(I_3,\theta_3;I_2,\theta_2) = 
{-A_m I_3 \over 4 h_{1}^2 \nu} \cos(m (\theta_2 - \Omega_b t) \pm 2 \theta_3) 
+ ...
\end{equation}


We focus on the particular resonance that we expect
is important and ignore all the other terms.
\begin{eqnarray}
H(&I_2,I_3;\theta_2,\theta_3)  = 
\nu I_3 + {\lambda  \over 32  \nu^2}  I_3^2 
+ \Omega I_2  ~~~~~~~~~~~~ \nonumber \\
& +  I_3 
{-A_m \over 4 h_{1}^2 \nu} \cos(m (\theta_2 - \Omega_b t) - 2 \theta_3)  
+ ...
\end{eqnarray}

To go into a coordinate system with an angle 
that librates when a particle is trapped in the resonance
we do a canonical transformation with
the following generating function 
\begin{equation}
F_2(J_3,J_2;\theta_3,\theta_2) 
       = J_2 \theta_2
- {J_3\over 2} [m(\theta_2 - \Omega_b t) - 2\theta_3]
\end{equation}
so that our resonant angle and new momenta
\begin{eqnarray}
\label{res_angle}
 \phi &=& \theta_3 -  {m\over 2} (\theta_2 - \Omega_b t)   \\
I_2 &=&  J_2 - J_3 {m\over 2} \nonumber \\
I_3 &=&  J_3 \nonumber
\end{eqnarray}
and $\theta_2$ is unchanged.
Our generating function is time dependent;
${\partial F_2/\partial t} =  {m\over 2} \Omega_b J_3 $.
Following this canonical transformation our Hamiltonian becomes 
\begin{eqnarray}
H(&J_2,J_3;\theta_2,\phi)  =    
[\nu - {m \over 2}(\Omega -\Omega_b)] J_3 
+ {\lambda  \over 32  \nu^2}  J_3^2  
\nonumber \\
& +  J_3 
{-A_m \over 4 h_{1}^2 \nu} \cos(2 \phi ) + \Omega J_2 +
 ...
\end{eqnarray}
The motion in the two different degrees of freedom are now decoupled
so we can neglect $J_2$ and write the Hamiltonian in the simpler form
\begin{equation}
\label{H_simple}
H(J_3;\phi)  =  \nu' J_3 + a J_3^2 - \epsilon J_3 \cos(2 \phi)
\end{equation}
where 
\begin{eqnarray}
\nu' &\equiv& \nu - {m \over 2}(\Omega -\Omega_b) \\
a &\equiv& {\lambda  \over 32  \nu^2}  \nonumber \\
\epsilon &\equiv&{ A_m \over 4 h_{1}^2 \nu}  \nonumber
\end{eqnarray}

Equation(\ref{H_simple}) has resonant term proportional to $J_3$ 
instead of proportional $J_3^{1/2}$ as was the case for
the inner Lindblad
resonance, considered by \cite{cont75}, and for 
e-e' orbit-orbit resonances
commonly studied in Celestial mechanics  \citep{SolarSystemDynamics}.
However the Hamiltonian is similar to that of 
J+2:J orbital resonances studied by \citet{borderies}.
Orbits captured in the 2:1 vertical ILR  were seen 
in the numerical simulations of \cite{combes}
however they expected the resonance to be first order in $J_3^{1/2}$
rather than second order as we have found here.
By symmetry the gravitational potential 
when expanded should not contain terms proportional to $z$,
and so terms $\propto J_3^{1/2}$, unless we consider a
global instability such as the fire-hose bar buckling instability.
Indeed \citet{pfenniger91} found that a temporal $z$-symmetry breaking (due
to a collective effect) would make the resonance first order.
\citet{friedli90} reported that N-body simulations which forced $z$-symmetry
did grow boxy/peanut shaped bulges but more slowly than
those which did not force the symmetry.

Unless the galaxy is extremely thin and the bulge 
small we expect that the 2:1 vertical ILR will
occur in the inner region of the bar (\citealt{pfenniger91,combes}).
It is also possible that the 1:1 vertical ILR will be present.
The 2:1 vertical ILR  produces banana-shaped orbits,  referred
to as BAN in the characteristic diagrams of \citet{pfenniger91}
and the 1:1 vertical ILR produces titled orbits that are referred
to as anomalous (ANO) by \citet{pfenniger91}.
If the galaxy is very thin it is also possible that
there will be orbits associated with the 
4:1 vertical ILR \citep{pfenniger85}.

If we concentrate on the 2:1 vertical ILR
then the perturbation term from the bar that is 
relevant is the $m=4$ Fourier term; if we focus
on the 1:1 vertical ILR then the $m=2$ term is relevant.
It has long been known that bars exhibit significant
$m=4$ Fourier components and contain boxy or square contours
when projected onto the plane of the galaxy \citep{atha90}.
From Figure 4 by \cite{n4314} we estimate that the $m=4$ Fourier component of
the barred galaxy NGC 4314 the potential peaks at 
$A_4 \sim -0.04$ in units of $v_c^2$ 
and at $A_2 \sim -0.16$ in the same units.
We can use these as typical values for the potential perturbations
in a barred galaxy.

{
\centering
}
\begin{quote}
\baselineskip3pt
{\footnotesize Fig.~1-- 
Contour levels of the Hamiltonian in Equation(\ref{ham_j3}) 
have been plotted where $x = \sqrt{ 2 j_3} \cos{\phi}$
and $y=\sqrt{ 2 j_3} \sin{\phi}$.
Only one resonance region exists
for ${\bar\nu \over \bar\epsilon }>1$, two exist
for $-1 < {\bar\nu \over \bar\epsilon } <1$ and three for   
${\bar\nu \over \bar\epsilon }<-1$.
The resonance bifurcates
at ${\bar\nu \over \bar\epsilon} = \pm1$ giving banana-shaped
orbits at $\phi = 0$ and upside down banana-shaped orbits at
$\phi = \pi$. 
The bar strength sets $\bar\epsilon$
and $\bar\nu$ varies with radius from the center of the resonance
where $\bar\nu = 0$.
Between ${\bar\nu \over \bar\epsilon} =-1 \rightarrow 1 $
no small $j_3$ circulating orbits exist.  We can think
of the progression of resonance sizes between
${\bar\nu \over \bar\epsilon} \sim -1 \rightarrow 1 $
as describing the radial variation of the width of the peanut.
We expect maximum vertical amplitudes at ${\bar\nu \over \epsilon} \sim -1$
which is on the outer, larger radius side of the resonance.
As the bar grows phase space moves downwards from plot to plot and
particles are trapped and then lifted upwards.  
Likewise if the bar slows down the same thing will happen.
}
\end{quote}
\smallskip

We now transform Equation (\ref{H_simple})
so that our coefficients are unitless.   
We put length in units of $r_{res}$, 
the center of the resonance where $\nu' = 0$ and 
time in units of $1/\nu$.   
Our action variables are in units of $r_{res}^2 \nu$ so we define
$j_3 \equiv  J_3 /r_{res}^2 \nu$.
We divide the entire Hamiltonian by $r^2_{res} \nu^2$
so that our new Hamiltonian becomes
\begin{equation}
\label{ham_j3}
h(j_3,\phi)  /(a r_{res}^2) = 
j_3^2  + \bar\nu j_3 -   \bar\epsilon j_3 \cos(2 \phi) 
\end{equation}
where
\begin{eqnarray}
\bar\nu &\equiv& {32 \nu' \nu \over r_{res}^2 \lambda }
 \sim  {- 32 \nu' \over \nu} { h_1^2 \over r_{res}^2 }             \\
\bar \epsilon &\equiv&{ 8 A_m  \over r_{res}^2  h_1^2 \lambda}  
 \sim {-8 A_m  \over v_c^2} { v_c^2 \over \nu^2 r_{res}^2}
\nonumber
\end{eqnarray}
and we have assumed that the 
vertical form of the potential can be approximated by 
$\lambda \sim {-\nu^2 \over h_1^2}$.
$\lambda$ should be negative because the vertical
oscillations are expected to be slower for particles 
traveling well above the 
galactic plane.

For a bar with a peak ${A_4 \over v_c^2 } \sim -0.04$, as estimated from
NGC 4314. $\nu r_{res} \sim 2 \Omega r_{res} \sim 2 v_c$ at the 
the 2:1 resonance,  and $\nu r_{res} \sim v_c$ at the
1:1 resonance.
We estimate that $\bar\epsilon$ peaks at $ \sim 0.08$ for both
resonances because $A_2 \sim 4 A_4$.
Barred galaxies have an 
observed ratio of bar length to peanut length 
of $2.7 \pm 0.3$ \citep{lutticke}.
Measurements of the ratio of bar length to thickness 
give a typical value of $\sim 10$ \citep{wilkinson} 
with the measurement of \citet{lutticke}
being on the high end at $14 \pm 4$, and the Milky Way bar
likely to be lower than the typical value.
Using these typical values we estimate that $ h_1/r_{res} \sim 0.3$.
\begin{eqnarray}
\bar\nu &\sim & -3 { \nu' \over \nu} 
       \left({ h_1/ r_{res}\over 0.3 }\right)^2\\
\bar \epsilon &\sim & 0.08 
       \left({- A_m/v_c^2 \over 0.04 }\right) 
       \left({2 \over  \nu r_{res}/v_c}\right)^2.  \nonumber
\end{eqnarray}

We expect the resonance to be important in the region
where $\bar\nu  = 0 \rightarrow \bar\epsilon$
or across a range $dr$;
\begin{equation}
{dr \ \nu'_{,r} \over \nu}={-A_m \over 4 \nu^2 h_1^2 }
\end{equation}
where $_{,r}$ denotes a derivative with respect to radius.
If $\bar\nu$ and $\nu'$ varies slowly with radius then 
the resonance can be felt over a significant range
of radius. 

\smallskip 
{
\centering
\includegraphics[angle=0,width=3.7in]{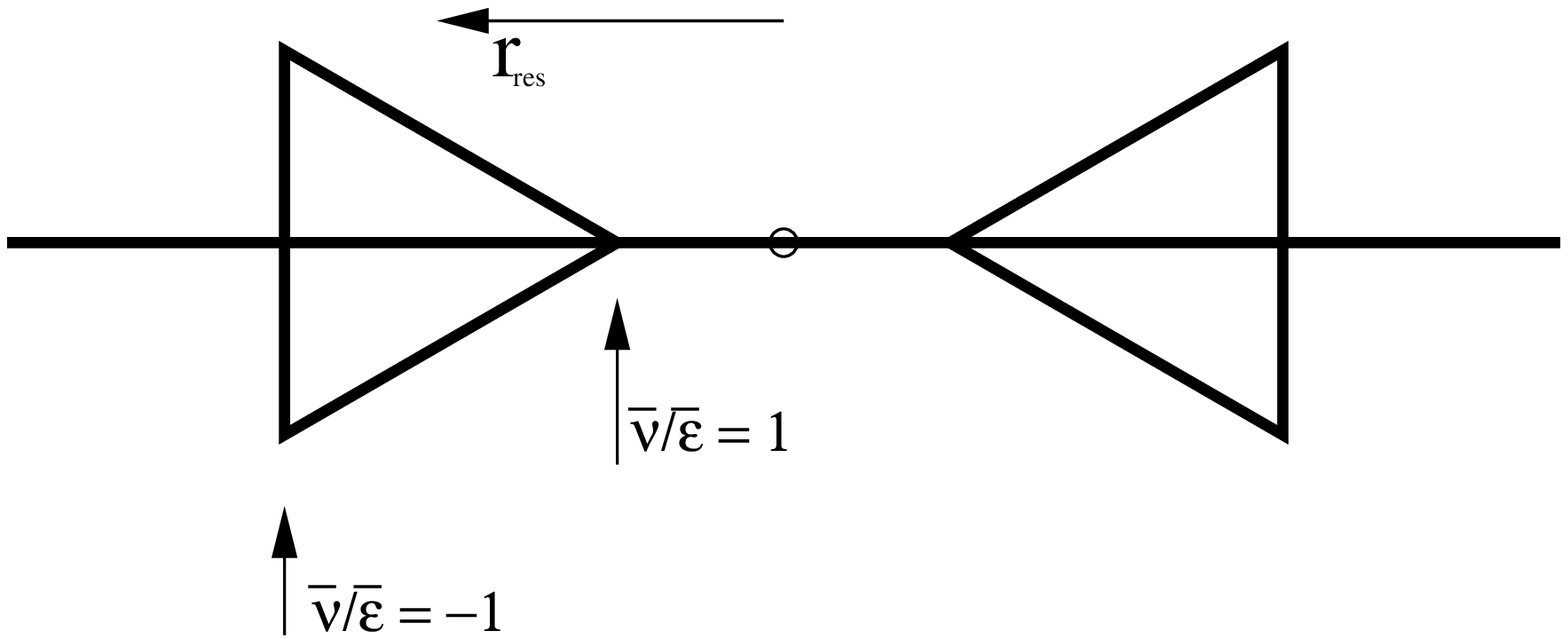}
}
\begin{quote}
\baselineskip3pt
{\footnotesize Fig.~2--
A cartoon of the edge-on galaxy labeling the approximate location of the peanut 
in terms of the values of ${\bar\nu\over\bar\epsilon}$.
}
\end{quote}

{
\centering
\includegraphics[angle=0,width=3.8in]{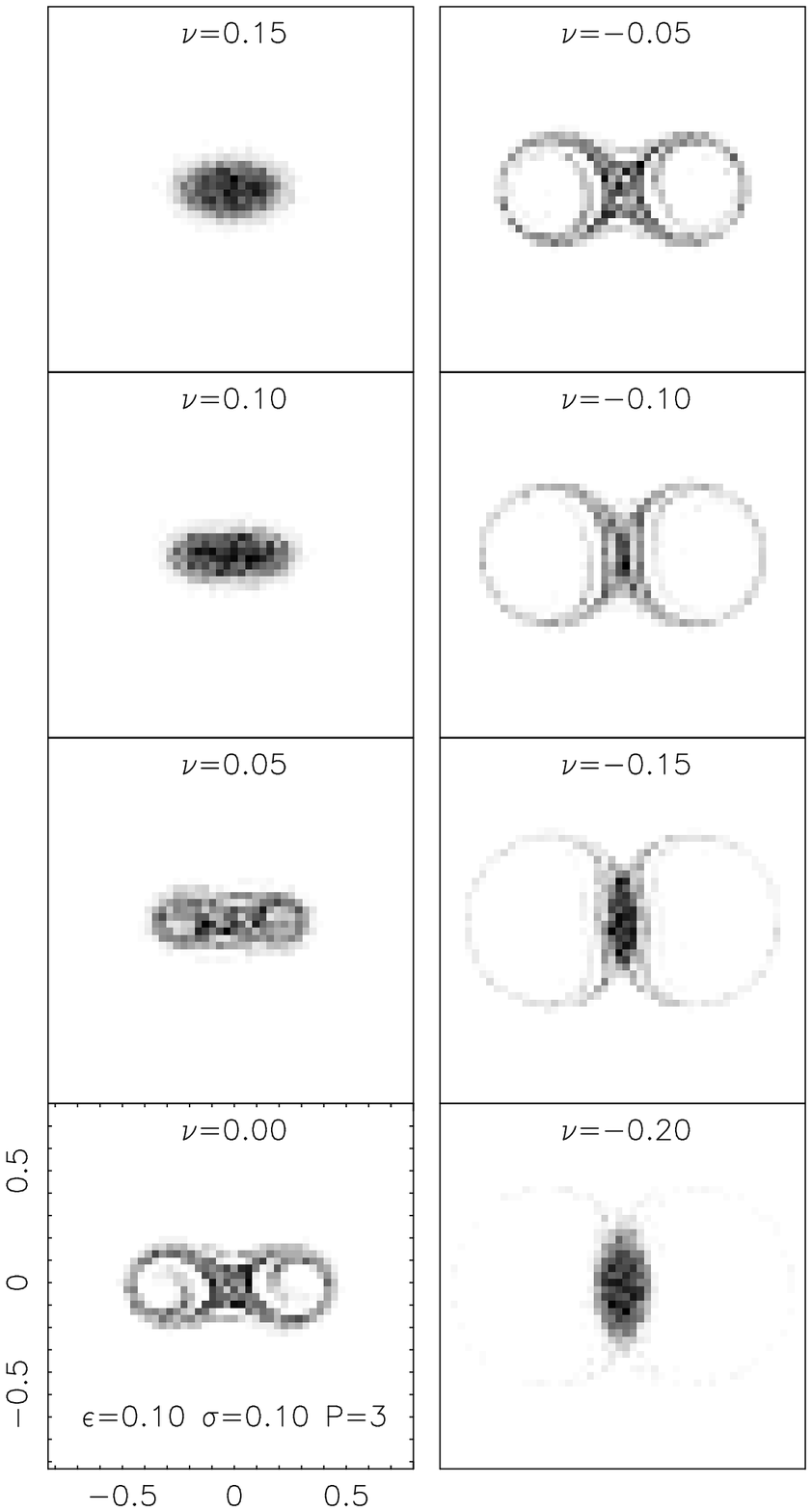}
}
\begin{quote}
\baselineskip3pt
{\footnotesize Fig.~3--
Particle distribution following growth
of a perturbation with $\bar\epsilon = 0.1$
for different values of $\bar\nu$.  The initial dispersion in $j_3$
was $\sigma = 0.1$.
The perturbation was grown in $P=3$ periods and then the integration
continued for 5 times longer than the growth period so that
orbits become more evenly distributed in phase.
$x$ and $y$ axes are defined as in Fig.~1.
}
\end{quote}

Given the definition of our variables we estimate that 
the expectation value 
\begin{equation}
\langle z^2\rangle \sim \langle j_3\rangle r_{res}^2.
\end{equation}
From Figure 1 we infer that when ${\bar\nu \over \bar\epsilon} = -1$
the minimum value of vertical oscillations occurs for the fixed
points at $j_3 =1$.
We therefore expect the maximum height of the 
the peanut to be at 
${\bar\nu \over \bar\epsilon} = -1$ where
$<j_3> \sim \bar\epsilon$.
For $\bar\epsilon \sim 0.08$  
this implies a peanut width of order $2 \sqrt{\bar\epsilon}r_{res} \sim 0.6 r_{res}$
where we gain the factor of 2 when we consider orbits of
both signs ($\phi=0$ and $\phi=\pi$).
This is sufficiently large to explain the observed vertical
extent of boxy/peanut shaped bulges.

\subsection{Structure of the resonance}

As shown by \citet{borderies} the resonance contains
3 separate regions.   To show the separate regions
we use a coordinate system $x = \sqrt{2 j_3}\cos{\phi}$ and
$y = \sqrt{2 j_3}\sin{\phi}$.  Contours of constant
$h$ (Eqn.~\ref{ham_j3}) are shown in Figure 1.
For ${\bar\nu\over \bar\epsilon} <-1$
three fixed points exist, for $-1<{\bar\nu\over\bar\epsilon}<1$ two fixed
points exist and for  ${\bar\nu\over \bar\epsilon}>1$ only one
fixed point exists.
Orbits circulating around
$\phi = 0$ correspond to banana-shaped orbits with
positive $z$ at the ends of the bar, and orbits 
correspond to banana-shaped orbits with negative $z$
at the ends of the bar.
As shown in Figure 8b of \citet{pfenniger91} banana
shaped orbits appear on the characteristic diagram
at a particular value of the Hamiltonian, consistent
with them appearing at a particular value of $\bar\nu\over \bar\epsilon$.
For orbits circulating around the origin, the resonant
angle, $\phi$, circulates; referred to as oscillation 
as opposed to libration in analogy with a pendulum.

We consider the signs of each term in Eqns.(\ref{H_simple},\ref{ham_j3}). 
$\nu>0$ but we expect that $\lambda, A_4, A_2 <0$.
Consequently $a<0$ and $\epsilon,\bar\epsilon >0$ and
$\bar\nu$ has the opposite sign of $\nu'$.
Because $\nu$ and $\Omega$ increase at small radii 
we expect $\nu'>0$ and $\bar\nu<0$ for $r > r_{res}$.
We see from Fig.1 that near $\bar\nu= -\bar \epsilon$ no orbits exist
with low values of $j_3$. This implies that all orbits
will rise out of the plane of the galaxy.
As $\bar\nu$ decreases the mininum size
of the vertical oscillations increases until ${\bar\nu \over \bar\epsilon} < -1$,
then orbits with small amplitude vertical oscillations appear 
again.  The vertical extent of the galaxy should increase until
a particular radius and then drop abruptly (see Figure 2).   
This would naturally correspond to a bowtie or peanut shape when viewed edge-on.

\subsection{Growing the bar}

To explore the simple model given in Eqn.(\ref{ham_j3})
we integrate particle trajectories assuming that the 
perturbation is growing.   In other words we assume
that $\bar\epsilon $ begins at zero and reaches
a maximum value a few rotation periods later.
We begin the integrations with a particle distribution
$p(j_3) \propto \exp(-j_3^2/ (2 \sigma^2))$
and with angle $\phi$ randomly chosen
between 0 and $2 \pi$.  4000 particles
were integrated using Eqn.(\ref{ham_j3}) for the simulations 
shown in Figures 3--6.
The growth time for the perturbation is given in
periods where $P = {2 \pi \over \nu}|a| r_{res}^2$.
Since $|a| r_{res}^2 \sim {r_{res}^2 \over h_1^2 32} \sim 1/3$,
$P$ is about 1/3 the bar growth time 
in units of the vertical oscillation period.

In the adiabatic limit 
the capture probability of a region of phase space depends
upon the ratio between the rate that phase space volume is swept across
the separatrix bounding 
this region divided by the rate that phase space volume is
swept across the entire separatrix bounding 
the entire resonance \citep{henrard,borderies}.
For $\bar\nu<0$ phase space initially looks like the panel 
on the bottom right of Fig.~1 and then moves upward on this figure.
When the separatrix touches the particle distribution centered
at the origin, particles
are captured into  librating regions near $\phi=0$ and $\pi$
(see Figures 3--4).
For $\bar\nu>0$ phase space initially looks like the panel on 
the top left of Figure 1 and then moves downward on this figure.
When $\bar\nu \sim  \bar\epsilon$ particles at low $j_3$ are captured
into the librating regions around $\phi=0,\pi$.
Figures 3--4 illustrate that
smoother particle distributions are achieved when
the growth rate is faster (shorter number of periods $P$) as we
would expect when the variation is no longer adiabatic, 
and when the initial momentum distribution is wider (larger $\sigma$).

Given a distribution in $j_3$ and $\phi$ resulting from the growth
of the perturbation we can predict the vertical distribution
as a function of azimuthal angle in the  plane of the galaxy.
For a value of the resonant angle $\phi$
and an azimuthal angle $\theta - \Omega_b t$ in the plane of the galaxy
in the frame rotating with the bar, $\theta_3$ is determined
via Eqn.(\ref{res_angle}) and $z$ is then determined
from the definition of our action angle variables (Eqn.~\ref{zdef}).
In Figure 5 we show vertical distributions for the action angle
distributions shown in Figures 3 and 4b for $m=2$ and 
in Figure 6 we show the equivalent but for $m=4$.

We see from Figures 5 and 6 that the vertical height
reaches a maximum near ${\bar\nu\over \bar \epsilon} \sim -1$
as we expected from the form of the resonance.
The height increases slowly between 
${\bar\nu\over \bar \epsilon} \sim 1$
to $-1$ and then drops swiftly, confirming conceptually
our bowtie picture shown in Fig.~2.
For $m=2$ (shown in Figure 5) corresponding to the 1:1
vertical ILR, the major axis of the bar
becomes bulbous, as would be expected from a sum 
of anomalous orbits of both orientations (shown in Figure 7). 
For $m=4$ (shown in Figure 5) corresponding to the 2:1
vertical ILR, both the major and minor axes of the bar
have increased vertical heights.  Again
this is expected from a sum of banana-shaped
orbits of both orientations (see Figure 7).

Observations of peanut shaped galaxies where both
the bar and the peanut are observed (e.g., \citealt{quillen})
suggest that the ends of the peanut are aligned with 
the bar.  So it could be that the 1:1 resonance orbit family is more
likely to explain the bulbous ends of boxy/peanut shaped bulges.
However when the radial oscillations
of the galaxy are taken into account the
banana-shaped orbits may still be viable, particularly
if the extent of their vertical oscillation is smaller
along the minor axis than it is along the major axis.

\smallskip
{
\centering
\includegraphics[angle=0,width=2.8in]{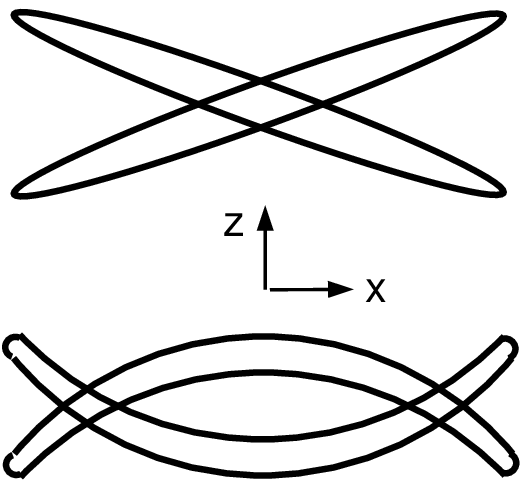}
}
\begin{quote}
\baselineskip3pt
{\footnotesize Fig.~7-- Anomalous orbits  at the 
1:1 resonance are shown on the top and give increased
widths at the ends of the bar.  Banana shaped orbits 
at the 2:1 resonance are shown on the bottom and 
give increased widths at the ends and sides of the bar.
The $x$ axis is assumed to lie along the major axis of the bar.
}
\end{quote}

We remind the reader that
in Section 2.1 we dropped the last term in Eqn.(\ref{Hz_can})
which is proportional to $I_1 I_3$.
As a bar grows the epicyclic amplitude $I_1$
of the stars will grow.
This has the effect of shifting the location
of the vertical resonance so that $\nu'$ becomes 
\begin{equation}
\nu' = \nu - {m\over 2}(\Omega - \Omega_b) 
+ (\kappa^2 - 3\Omega^2){\nu I_1\over 4 \kappa}.
\end{equation}
For a flat rotation curve $\kappa^2 = 2 \Omega^2$
so we expect that $\nu'$ will be more negative than expected 
if this additional term is neglected.
This has the affect of moving the resonance outward in radius.
One reason that the 2:1 ILR resonance was considered a logical
choice for exciting the peanut is that when the galaxy
is thin the 1:1 may only occur at extremely small radii, 
well inside the galaxy bulge.  However the estimates for the resonance
locations were made assuming $I_1\sim 0$
(e.g., \citealt{pfenniger}).   The point here is
that the growth of the epicyclic amplitude will shift
the location of the vertical resonances outwards so
that the lower order resonances may manifest at larger
radii than previously considered.
It is possible that the growth of the bar causes the vertical
resonances to sweep through the inner part of the galaxy.

\section{Summary and Discussion}

We have presented a simple Hamiltonian dynamical model for
the vertical inner Lindblad resonances
in the presence of a bar perturbation.
The Hamiltonian model is similar to that introduced
by \citet{cont75} to describe the 2:1 Lindblad resonances
in the plane of the galaxy
but uses the vertical action angle variables.
Using this model we consider the 
growth of a boxy/peanut  shaped bulge via
resonance trapping in the vertical inner Lindblad resonances
caused by the growth of the bar.
Following the growth of the bar,
our model predicts that the height of the peanut
increases as a function of radius until it reaches
a maximum at the outermost radius which is set by the 
condition ${\bar\nu \over \bar\epsilon} = -1$.
Our model naturally accounts for the sharp outer
edge of observed boxy/peanut shaped bulges.
The maximum total vertical width of the peanut 
is of order $\sim 2 \sqrt{\bar\epsilon} r_{res}$ which we estimate
for a typical bar could be about two thirds $r_{res}$,
where $r_{res}$ is the radial location of the resonance.
This model is therefore capable of accounting for the observed
widths of boxy/peanut shaped bulges.

The scenario explored here suggests that the boxy/peanut
shaped bulge is grown at the same time as the bar forms.
We predict that recently formed bars should
manifest strong boxy/peanut shaped bulges and barred
galaxies should never be found without boxy/peanut shaped
bulges.    
If the boxy/peanut shapes take nearly a Hubble time to grow, as
found in the N-body simulations of \cite{combes}, then
there should be barred galaxies which
lack boxy/peanut shaped bulges.
However \cite{bureau} considered a control
sample of edge-on galaxies lacking boxy/peanut bulges.
No examples of barred galaxies that lack a boxy/peanut
shaped bulge have yet been identified.
\citet{pfenniger91} found that the boxy/peanut only required
3-4 bar growth periods to grow and this could be significantly
shorter, only a few Gyrs.  If this timescale were 
more appropriate then the lack of barred galaxies which do 
not display boxy/peanut shaped bulges is less of a problem.

One way to probe the formation timescale of the boxy/peanut
is to search for barred galaxies which have recently
formed bars.  If boxy/peanuts  require many rotation periods
to form then these barred galaxies should have normal (non-boxy) bulges.
Because of the high rate of star formation  along the bar
\cite{quillen} inferred that NGC 7582
was a case of a recently formed bar.
They pointed out that the galaxy bulge did show a strong prominent peanut
and so argued that boxy/peanut shaped bulges are
formed quickly after the onset of a bar.
Observations of edge-on galaxies are therefore consistent with
a rapid boxy/peanut formation mechanism such as we have proposed here.  

Our scenario does not require the galaxy to buckle
in order to form the boxy/peanut shaped bulge.
The fire-hose instability scenario for bar-buckling
(beautifully illustrated in 3D N-body simulations
by \citealt{raha}) predicts that the galaxy will
literally be bent in a U-shape
while the galaxy buckles, however
no examples of U-shaped edge-on galaxies have been
identified in galaxy surveys (with the exception
of galaxies which are believed to be ram pressure stripped in clusters).
It is likely that the period of time during which
the galaxy is U-shaped is short.  However
galaxy collisions have a similar estimated lifetime and
are detected in nearby galaxy surveys.
We note that \citet{friedli90} reported that N-body simulations
with fixed $z$-symmetry did grow boxy/peanut shaped bulges, 
though at a slower rate than unconstrained simulations.


We have constructed an analytical framework which
can be used to predict
the height distribution of the peanut as a function 
of radius and the azimuthal angle
in the galaxy plane.  The nature of the dependence
is primarily determined by the resonance responsible
for the vertical excitations.
A 3 dimensional study of a nearby boxy bulge 
(such as that seen in our galaxy) could determine which
resonance is most likely to be responsible.
The shape of the boxy/peanut shaped bulge is also sensitive 
to the dependence of the gravitational potential on height 
from the mid-plane and on the way the bar grew.  
Further study of boxy/peanut
shaped bulges may allow us to place constraints
on the vertical shape of the galaxy and on 
the early evolution of the bar.

The toy model introduced in this paper was designed 
to illustrate the process of resonance {\bf capture} during the growth 
of a bar.  We have argued that the growth of the bar itself
causes stars to become trapped into vertical
resonances and so lifted out of the plane of the galaxy.
However, simultaneous growth of a bar and boxy/peanut
shaped bulge has not yet been observed in three
dimensional N-body simulations of non-bending bars.
This could in part be because N-body simulations are
are typically thicker than thin bars such as NGC 7582.
Moreover, N-body simulations \citep{pfenniger91,friedli90,combes} 
have shown that the boxy/peanut continues to grow well after
the bar is stable.    Since resonance capture is likely to be
a natural consequence of bar growth, it might be useful
to reexamine N-body simulations and verify if orbits
are trapped into vertical resonances during the bar
growth period.  After the bar stabilizes,
the analytical framework developed here 
might also be applied toward predicting the evolution of stellar orbits 
caused by secular evolution of a bar. 
For example a slow decrease in pattern speed will also trap 
particles in the vertical resonances and cause them to be
lifted into higher orbits. In this case $\bar\nu$ decreases
with time so that phase space  
moves downwards in Figure 1, causing similar
phenomenolgy as growing the bar; particles trapped into
resonances can be lifted higher as the bar evolves.


\begin{figure*}
\vspace{18.0cm}
\figurenum{4}
\caption[]{
a) Similar to Fig.~3 except 
the initial dispersion in $j_3$ is $\sigma = 0.05$.
b) Similar to a) except the perturbation was
grown in 10 rotation periods.
}
\end{figure*}

\begin{figure*}
\vspace{20.0cm}
\figurenum{5}
\caption[]{
a) For the distribution in action angle variables displayed in Fig.~3
this shows the vertical distribution as a function 
of azimuthal angle for the case $m=2$ corresponding to the 1:1 resonance.  
The $x$ axis shows the azimuthal angle, $\theta - \Omega_b t$,
in the plane of the galaxy in the frame that rotates with the bar
and is given in degrees. 
The $y$ axis shows the $z$ distribution in units of $r_{res}$. 
b) Same as a) but for $P=10$ (also shown in Figure 4b).
We see from these figures that we expect a maximum
peanut width at the radius where ${\bar\nu\over\bar\epsilon} = -1$.
}
\end{figure*}

\begin{figure*}
\vspace{20.0cm}
\figurenum{6}
\caption[]{
Same as Fig.~5a,b but for the case when $m=4$.
}
\end{figure*}

\acknowledgments                             
This work would not have been carried out without helpful
discussions with Andrew Anissi and Larry Helfer.
%
%
%
A.~C.~Q.~gratefully thanks the Technion for hospitality and support
during the fall of 2001.

\appendix
\setcounter{equation}{22}


To put the Hamiltonian into
action angle variables to first order 
we use the generating function
\begin{equation}
F_1(z,\theta') = - {\nu \over 2 } z^2 \tan{\theta'}
\end{equation}
Following this canonical transformation the $z$ part of the Hamiltonian
restricted to $r=r_c$
\begin{equation}
H_z(p_z;z)  = {1\over 2}(p_z^2 + \nu^2 z^2) + {\lambda z^4 \over 4!} 
\end{equation}
becomes
\begin{equation}
H_z(I';\theta')  = \nu I_3' + {\lambda I'^2  \over 6 \nu^2} \cos^4(\theta')
\end{equation}
To put the Hamiltonian into a form that doesn't depend on the angle
we must do another canonical transformation.
Note that $\cos^4{\theta} = { 1\over 8} \cos{ 4 \theta} 
+ {1 \over 2} \cos{ 2 \theta} + {3\over 8}$.

We try the following generating function
\begin{equation}
F_2(\theta',I_3) = I_3 - {\lambda \over 6 \nu^2} I_3^2 ( {1\over 4} \sin{2 \theta'} + {1\over 32} \sin{4 \theta'})
\end{equation}
which leads to new variables $I_3, \theta_3$ such that
\begin{eqnarray}
I' &=&  I_3 + { \lambda \over 6 \nu } I_3^2 
    \left( {1 \over 2} \cos{2 \theta'} + {1\over 8} \sin {4 \theta'}\right) \\
\theta_3 &=&  \theta' - {\lambda \over 6 \nu^2} I_3'
     \left({1 \over 4} \sin{2 \theta'}  + {1\over 32} \sin{4 \theta'} \right) \nonumber
\end{eqnarray}
To fourth order in $I_3^{1/2}$ the Hamiltonian becomes 
\begin{equation}
H_z(I_3;\theta_3)  = \nu I_3 + {\lambda I_3^2  \over 32  \nu^2}
\end{equation}
which is independent of $\theta_3$ so that $I_3$ is a conserved quantity
and we find that we have successfully transfered the Hamiltonian
into action angle variables.

Via a similar canonical transformation the term
proportional  to $(r-r_c ) z^2$ in Equation (\ref{Hz}) disappears
but the term proportional to $(r - r_c)^2 z^2$ remains yielding
the additional term 
$ (\kappa^2 - 3\Omega^2) {\nu I_1 I_3 \over 4\kappa }$
which we have included in Eqn.~(\ref{Hz_can}).

  
{}

\end{document}